\documentclass{article}
\begin{document}  
\title{Singularity-Free Cylindrical Cosmological Model}  
\author{L. Fern\'andez-Jambrina \footnote{Permanent address: 
Departamento de Geometr\'{\i}a y Topolog\'{\i}a, 
Facultad de Ciencias Matem\'aticas, 
Universidad Complutense de Madrid, 
E-28040-Madrid, Spain}  \\ Theoretisch-Physikalisches Institut\\Max-Wien-Platz 
1\\Friedrich-Schiller-Universit\"at-Jena\\07743-Jena, Germany}
 
\maketitle
\begin{abstract} 
A cylindrically symmetric perfect fluid spacetime with no curvature
singularity is shown. The equation of state for the  perfect fluid is that of
a stiff fluid. The metric is  diagonal and non-separable in comoving
coordinates for  the fluid. It is proven that the spacetime is
geodesically complete and globally hyperbolic.\end{abstract}
\noindent PACS {04.20.Jb, 98.80.Hw}

\section{Introduction}
Due to the powerful singularity theorems (cfr. for instance 
\cite{HE}) it was 
widely believed that 
cosmological models were to have an initial singularity. However 
in 
1990 
Senovilla \cite{Seno} showed the first cosmological perfect fluid 
solution of the Einstein equations with regular scalar curvature 
invariants. It corresponds to a cylindrically symmetric spacetime 
filled with an isotropic radiation perfect fluid. In \cite{Chinea} 
it 
was shown that this spacetime is indeed singularity-free. This is
in 
full accordance with the singularity theorems, as it is to be 
expected, since although it fulfills the energy, generic and causality 
conditions, it does not have have any of the causally trapped 
sets 
required by the theorems (closed trapped surfaces, compact 
achronous 
sets without edge,...). This special solution was generalized in a 
subsequent paper \cite{Ruiz} where an Ansatz of separability of 
variables for diagonal orthogonally transitive commuting $G_2$
metrics 
was extensively explored. A thorough discussion of this Ansatz can 
be 
found in \cite{Esc}. This family is shown to be included in a wider class of
separable cosmological models which comprises FLRW universes \cite{family}. Other
properties of these solutions, such as their inflationary behaviour,  their
generalized Hubble law or the feasability of constructing a realistic
non-singular cosmological model, are studied therein. Other non-singular
cylindrically symmetric  perfect fluid spacetimes have been found \cite{Diag}, 
\cite{Jerry}. Together with some solutions in \cite{Mars}, these are the only
non-diagonal ones that are known.  Furthermore, in the family of solutions
included in \cite{Sep},  there  are non-separable non-singular perfect fluid
spacetimes, with and without symmetry axis \cite{Kramer}. 

In this paper we present a cylindrically symmetric stiff perfect fluid 
solution which is non-separable in comoving coordinates and has 
non-singular scalar curvature invariants. Due to these facts it is 
not included in the families quoted before. The physical properties of
 these metrics and their relation to other families of solutions are
investigated.

\section{The metric}
The line element for a spacetime that admits an Abelian two-dimensional
orthogonally transitive group of isometries acting on
spacelike surfaces can be cast in the form 
 \begin{equation}\label{metric}
ds^2={\rm e}^{ K(t,r)}\,(-dt^2+dr^2)+{\rm e}^{-U(t,r)}\,dz^2+
 {\rm e}^{U(t,r)}\,r^2\,d\phi^2,
\end{equation}
where $\phi$ and $z$ are coordinates adapted to the commuting generators (in
this chart, $\xi=\partial_\phi$, $\zeta=\partial_z$) of the group. The
remaining coordinates, $\{t,r\}$, have been chosen so that the
line element induced in the subspaces orthogonal to the Killing
orbits is isotropic in these coordinates.

Following \cite{Kramer}, if we are to have a regular symmetry axis on
the locus where $\Delta=g(\xi,\xi)=0$, then we have to impose that
\begin{equation}
\frac{g({\rm grad}\,\Delta,{\rm grad}\,\Delta)}{4\,\Delta}\rightarrow
1,\label{reg} \end{equation}
on approaching the axis. 

The metric functions for the spacetime that we want to show are given in this
set of coordinates by

\begin{equation}
K(t,r)=\frac{1}{2}\,{ \beta}^{2}\,{r}^{4 } + ( \alpha+\beta)\,{r}^{2} +
2\,{t}^{2}\,{ \beta}
 + 4\,{t}^{2}\,{ \beta}^{2}\,{r}^{2}
 \end{equation}
 \begin{equation}
 U(t,r)={ \beta}\,(\,{r}^{2} + 2\,{t}^{2}\,)
 \end{equation}
\begin{equation} -\infty<t,z<\infty, \ \ \ \  0<r<\infty,\ \ \ \ 0<\phi<2\,\pi
\end{equation}
and are easily checked to satisfy the regularity condition (\ref{reg}) on the set
of events defined by $r=0$. There are no further isometries than the ones
that have been implemented from the beginning and therefore this spacetime is
properly said to admit cylindrical symmetry. The only restriction that we impose
on the parameters $\alpha$ and $\beta$ is that both of them  are positive.

 The matter content in this spacetime is a stiff
perfect fluid, whose density and pressure, 

\begin{equation}
\mu=p= \alpha\,{\rm e}^{-K(t,r)},
\end{equation}
 are regular everywhere in this chart. When $\alpha$ equals zero we have a
vacuum solution, from which the stiff fluid metric can be recovered by means
of the Wainwright, Ince, Marshman algorithm \cite{Wain}. However this
solution has not been obtained by this method, although it could be used as it
is done in \cite{Jerry} to generate further models with the same vacuum metric.

The velocity of the fluid takes
the form, \begin{equation} {u}= {\rm e}^{ -\frac{1}{2}\,K(t,r)}\partial_t,
 \end{equation}
and therefore the coordinates are comoving. From the expression for the line
element (\ref{metric}) it is clear that it cannot be rendered separable in
comoving coordinates.

The acceleration of the fluid has  projection only on the
radial direction,
 \begin{equation}
a= {r}\, \left( \! \,{ \beta}^{2}\,{r
}^{2} + { \alpha} + { \beta} + 4\,{ \beta}^{2}\,{t}^{2}\, \! 
 \right) \, \partial_r,
\end{equation}
due to the orthogonal
transitivity requirement and that the fact that the velocity is orthogonal to the
orbits of the group of isometries. For the same reason the vorticity is zero. 

In an orthonormal coframe $\{\theta^0,\theta^1,\theta^2,\theta^3\}$, where the
four independent differential one-forms take the form,

\begin{equation}
\theta^0={\rm e}^{\frac{1}{2}\,K(t,r)}\,dt,\ \ \ \ \theta^1={\rm
e}^{\frac{1}{2}\,K(t,r)}\,dr, \ \ \ \ \theta^2={\rm e}^{-\frac{1}{2}
U(t,r)}\,dz, \ \ \ \ \theta^3=
 {\rm e}^{\frac{1}{2}U(t,r)}\,r\,d\phi,\label{coframe}
\end{equation}
the shear tensor constructed with the derivatives of the velocity $u$ can be
written as,  
\begin{equation}
\sigma=\frac{4}{3}\,{ \beta
}\,{t}\,{\rm 
e}^{-\frac{1}{2}\,K(t,r)}\,\left\{(\,1 + 2\,{ 
\beta}\,{r}^{2}\,)\,\theta^1\otimes \theta^1-(\,2 + { 
\beta}\,{r}^{2}\,)\,\theta^2\otimes \theta^2+(1-\,{ \beta}\,{r}^{2
})\,\theta^3\otimes 
\theta^3\right\}.
\end{equation}

The expression for the expansion, $\Theta$,  of the cosmological fluid, 
\begin{equation}
{ \Theta}=2\,\beta\,t\,(\,1 + 2\,{ \beta}\,{r}^{2}\,)\,{\rm 
e}^{-\frac{1}{2}\,K(t,r)}, 
\end{equation}
allows us to calculate the deceleration parameter, $q$,  for this universe, 

\begin{equation}
{q} = 2 -  \frac {3}{2}\,\displaystyle 
\frac {1}{{ \beta}\,{t}^{2}\,(\,1 + 2\,{ \beta}\,{r}^{2}\,)},
\end{equation}
from
the action of the vector field $u$ on the inverse of $\Theta$, \cite{Ellis}
\begin{equation} u\left(\frac{1}{\Theta}\right)=\frac{1}{3}\,(1+q).
\end{equation}

Since $q$ is positive in the time span $(-t_{inf},t_{inf})$,
\begin{equation}
t_{inf}= \sqrt{\frac {3}{4\,{ \beta}\,(1 + 2\,{
 \beta}\,{r}^{2})}},
\end{equation}
we are led to conclude that this spacetime has an inflationary epoch, which is
longer the closer the observer is to the
symmetry axis.  Furthermore the scale factor of the universe can be defined
\cite{Ellis} as a solution of the differential equation,
\begin{equation} u(R)=\frac{\Theta}{3}\,R,
\end{equation}
which in our case can be solved, introducing an arbitrary function $C$ of the
spacelike coordinates, 
 \begin{equation}
{\rm R}(\,{t,r,z,\phi}\,)={\rm C}(r,z,\phi)\,{\rm e}^{\frac{1}{3}\,{
\beta}\,{t}^{2}\, (\,1 + 2\,{ \beta}\,{r}^{2}\,)}.
\end{equation}

\section{Regularity of the metric}

Since the scalar invariants that can be formed with the metric and the
Riemann curvature are polynomials of the
density and the pressure of the fluid and the components of the Weyl tensor, we
shall calculate the latter ones in order to show that the curvature scalars are
not singular. In the null tetrad that can be naturally constructed with the
one-forms of (\ref{coframe}), the components of the Weyl tensor can be shown to
be,

\begin{equation}
\Psi_0=\frac{1}{2}\left(f_1(t,r)+f_2(t,r)\right) \,{\rm e}^{ 
-K(t,r) }
\end{equation}

\begin{equation}
\Psi_1=0
\end{equation}
\begin{eqnarray}
{ \Psi_2}= {\displaystyle \frac {1}{6}}\, \left( \! \,  3\,{
 \beta}^{2}\,{r}^{2} + 3\,{ \beta} - { \alpha} - 12\,{ 
\beta}^{2}\,{t}^{2}\, \!  \right) \,{\rm e}^{ -K(t,r) }
\end{eqnarray}
\begin{equation}
{ \Psi_3}=0
\end{equation}
\begin{equation}
\Psi_4=\frac{1}{2}\left(f_1(t,r)-f_2(t,r)\right) \,{\rm e}^{ 
-K(t,r) },
\end{equation}

\begin{equation}
f_1(t,r)=- 3\,{ \beta}
 + 2\,{ \beta}^{3}\,{r}^{4} + 3\,{ \beta}^{2}\,{r}^{2} +  \alpha
\,(1+2\,\beta\,r^2) + 24\,{ \beta}^{3}\,{r}^{2}\,{t}^{2} +12\,{ 
\beta}^{2}\,{t}^{2}
\end{equation}
\begin{equation}
f_2(t,r)=12\,{ \beta}^{3}\,{r}^{3}\,{t} + 12\,{ 
\beta}^{2}\,{r}\,{t} + 16\,{ \beta}^{3}\,{t}
^{3}\,{r} + 4\,{ \alpha} \,{ \beta}\,{t}\,{r}.
\end{equation}

From the expressions for the components of the Weyl tensor and the density and
the pressure of the fluid it follows that they vanish when either $t$ or $r$ tend
to infinity. Hence the spacetime is flat for large values of the time and radial
coordinate. It is also easy to check that the Weyl components are regular
everywhere and so are the pressure and the density. Therefore all the curvature
invariants are regular. The spacetime has a low matter content for large
negative values of the time coordinate and undergoes a contracting epoch until
$t$ reaches the zero value. For positive values of $t$ this universe is
expanding. The change from contraction to expansion without developing a
curvature singularity appears also in other non-singular cosmological models
\cite{Seno}, \cite{family}, \cite{Diag}.

The gradient of the comoving time coordinate $t$ is always negative. It is
therefore a cosmic time \cite{HE} and the spacetime is causally stable. In
particular this implies weaker causality conditions suchs as the chronology
condition.

The strong and dominant energy conditions are satisfied since the density of
the fluid is positive everywhere and the equation of state corresponds to a
stiff fluid. Moreover, the
energy-momentum tensor does not vanish anywhere. This means that for every
non-spacelike vector {\bf $W$} the contraction with the Ricci tensor $R(W,W)$
is greater than zero and it implies that the generic condition is
satisfied as well \cite{Beem}.

There is a static limit for the metric which amounts to take the parameter
$\beta$ equal to zero. The resulting metric,

\begin{equation}
ds^2={\rm e}^{\alpha\,{r}^{2}}\,(-dt^2+dr^2)+dz^2+r^2\,d\phi^2,
\end{equation}
can be seen to be also the static limit of \cite{Diag}.

\section{Geodesic completeness}

In order to determine whether the spacetime is geodesically
complete, we have to study the equations for the causal geodesics,

\begin{equation}
\nabla_vv=0, \ \ \ v=\dot t\,\partial_t+\dot r\,\partial_r+\dot
z\,\partial_z+\dot \phi\,\partial_\phi, \end{equation}
 where $v$ is the tangent vector along the geodesic and the dot stands for the
derivative with respect to the affine parameter $\tau$. We shall introduce a
constant of motion $\delta$, which takes the zero value for null geodesics and
one for timelike geodesics, that is,

\begin{equation}\label{delta}
g(v,v)=-\delta.
\end{equation}

The existence of isometries simplifies the problem, since two new independent
constants of motion arise,

\begin{equation}
L=g(v,\partial_\phi)=r^2\,{\rm e}^{U(t,r)}\,\dot\phi,\ \ \ \
Z=g(v,\partial_z)={\rm e}^{-U(t,r)}\,\dot z, 
\end{equation}
 as a consequence of the geodesic and the Killing equation.

Since \ref{delta} can be cast in the form,

\begin{equation}
\dot t^2-\dot r^2={\rm e}^{- K(t,r)}\,\{\delta+Z^2\,{\rm e}^{U(t,r)}+
 L^2\,{\rm e}^{-U(t,r)}\,r^{-2}\},
\end{equation}
it is convenient to parametrize $\dot t$ and $\dot r$ with hyperbolic functions 
of a new variable, $\xi$, as it is done in \cite{Chinea}, in order to lower the order of the geodesic
equations. The final system of equations for future-directed geodesics
(past-directed ones can be handled in a similar way) is first order,

\begin{eqnarray}\label{system}\dot\xi&=&-
{\rm e}^{-\frac{1}{2}\,K(t,r)}\,\left\{\frac{r\,\cosh\xi\,(Z^{2}\,
A(t,r)+L^{2}\,B(t,r)+\delta \,C(t,r)
)}{\sqrt{G(t,r)}}\right.+\\ \nonumber&+&\left.\frac{t\,\sinh\xi\,(Z^{2}\,D(t,r)
+L^{2}\,E(t,r)+\delta\, F(t,r))}{\sqrt{G(t,r)}}\right\} \end{eqnarray} 

\begin{equation}
\dot t={\rm e}^{K(t,r)/2}\,G(t,r)\,\cosh\xi
\end{equation}

\begin{equation}
\dot r={\rm e}^{ K(t,r)/2}\,G(t,r)\,\sinh\xi
\end{equation}

\begin{equation}
\dot z={\rm e}^{U(t,r)}\,Z
\end{equation}

\begin{equation}
\dot\phi={\rm e}^{-U(t,r)}\,r^{-2}\,L,
\end{equation}
and the last two equations are just quadratures that can be integrated after
solving the first three equations. We have introduced seven functions of $t$ and
$r$,

 \begin{equation}
A(t,r)=(\beta^
{2}r^{2}+4\,\beta^{2}\,t^{2}+\alpha+2\,\beta)\,{e^{U(t,r)}}
\end{equation}
\begin{equation}
B(t,r)= (\beta^{2}-\frac{1}{r^4}+{\frac {
\alpha+4\,\beta^{2}t^{2}}{r^2}})\,{e^{-U(t,r)}} \end{equation}
\begin{equation}
C(t,r)= \beta^{2}r^{2}+\alpha+\beta+4\,\beta^{2}\,t^{2}
\end{equation}
\begin{equation}
D(t,r)= 4\,\beta\, (1+\beta\,r^{2})\,{e^{U(t,r)}}
\end{equation}
\begin{equation}
E(t,r)= 4\,\beta^{2}\,{e^{-U(t,r)}}
\end{equation}

\begin{equation}
F(t,r)= 2\,\beta\,\left (1+2\,\beta\,r^{2}\right )
\end{equation}

\begin{equation}
G(t,r)={\delta+Z^{2}{e^{U(t,r)}}+{\frac {L^{2}{e^{-U(t,r)}}}{r^{2}}}}.
\end{equation}

The only geodesics that fall out of this scheme are radial null geodesics,
$\delta=L=Z=0$. For them the analysis is fairly simple, since $\dot t=|\dot r|$ 
and the only equation that is left for integration is,

\begin{equation}
\ddot r+2\,r\,(\beta^2\,r^2+\alpha+\beta+4\,\beta^2\,t^2)\,\dot
r^2+4\,\beta\,t\,(1+2\,\beta\,r^2)\,\dot t\,\dot r=0, \end{equation}
which can be seen to have a first integral, $h$,

\begin{equation}
h=\dot r\,{\rm
e}^{2\,\beta\,t^2\,(1+2\,\beta\,r^2)+(\alpha+\beta)\,r^2+\frac{1}{2}\,\beta^2\,r^4}.
\end{equation}

The completeness of these geodesics is obvious, since $\dot r$
is confined finite in the interval $[-|h|,|h|]$.

Going back to the generic equations \ref{system}, we shall explore the
appearance of divergences that may occur when the radius, $r$ is either too
small or too large or when the time coordinate, $t$, is too large.

If the radius, $r$, grows too large, so that $r^4>\beta^{-2}$, $\dot r>0$
($\sinh\xi>0$), then $\dot\xi$ is negative when the time coordinate, $t$, is
positive. Therefore $\xi$ decreases and $r$ does not diverge for finite affine
parameter since $\dot r$ cannot grow arbitrarily large. If $t$ is negative,
since the time coordinate grows at least as fast as $r$, $t$ becomes positive
before $r$ diverges.

The terms with negative powers of $r$ do not yield singularities when the
geodesics approach the axis ($\dot r<0$, $\sinh \xi<0$). The reason is that
$\dot\xi$ becomes positive when the geodesic is close to $r=0$ and
prevents the radial coordinate from decreasing too quickly. This happens because
the only negative terms in the expression of
$\dot\xi$ are either bounded or overcome by the term in $r^{-4}$ in $B(t,r)$ for
decreasing radial coordinate. 

Also if $t$ grew too fast, this would mean that $r$ would grow or decrease
quickly and the previous reasonings would prevent divergences from appearing.

Hence we are led to conclude that every causal geodesic is complete and
therefore the spacetime is {\em geodesically complete}.

 From this analysis it follows
that every null geodesic intersects once and  only once every timelike
hypersurface $t=const.$ Therefore these hypersurfaces
 are Cauchy surfaces and the spacetime is globally hyperbolic \cite{Geroch}.

\section{Discussion}

It has been shown that the spacetime is singularity-free and globally
hyperbolic. Since energy, generic and causal conditions are fulfilled, it is
clear that this spacetime does not have causally trapped sets \cite{HE},
\cite{Beem}. The gradient of the transitivity surface element is spacelike as in
all known non-singular models.

The question of whether there is an open set of non-singular cosmological models
in a reasonable topology remains open. However the fact that both this spacetime
and the one shown in \cite{Diag} have the same static limit suggest that more
solutions could be found by suitably adding new parameters to the static
solution.

\vspace{0.3cm}
\noindent{\em The present work has been supported by Direcci\'on General de
Ense\~nanza Superior Project PB95-0371 and by a DAAD (Deutscher Akademischer
Austauschdienst) grant for foreign scientists.  The author wishes to thank
Prof. F. J. Chinea and Dr. L. M. Gonz\'alez-Romero for valuable discussions
and Prof. Dietrich Kramer and the Theoretisch-Physikalisches Institut of the 
Friedrich-Schiller-Universit\"at-Jena for their hospitality.}

\end{document}